\documentclass[superscriptaddress,twocolumn,
 amsmath,amssymb,
 aps,prl]
{revtex4-1}
\usepackage{amsmath,amssymb}

\usepackage{graphicx}
\usepackage{dcolumn}
\usepackage{bm}
\usepackage[bottom]{footmisc}
\usepackage{pdfpages}
\makeatletter

\usepackage[colorlinks=true,
    linkcolor=blue,
    citecolor=red,
    filecolor=magenta,      
    urlcolor=cyan,
    pdftitle={DiffusiveEntanglementGrowth},]{hyperref}

\AtBeginDocument{\let\LS@rot\@undefined}
\makeatother

\begin{document}

\title{Diffusive entanglement growth in a monitored harmonic chain}

\author{Thomas Young} 
 \affiliation{School of Physics and Astronomy, University of Birmingham, Edgbaston, Birmingham, B15 2TT, UK}
  \affiliation{Department of Physics, King’s College London, Strand WC2R 2LS, UK}

 \author{Dimitri M. Gangardt}
 \affiliation{School of Physics and Astronomy, University of Birmingham, Edgbaston, Birmingham, B15 2TT, UK}

\author{Curt von Keyserlingk} 
 \affiliation{Department of Physics, King’s College London, Strand WC2R 2LS, UK}

\date{\today}
\begin{abstract}
We study entanglement growth in a harmonic oscillator chain subjected to the weak measurement of observables which have been smeared-out over a length scale $R$. We find that entanglement grows diffusively ($S \sim t^{1/2}$) for a large class of initial Gaussian states provided the measurement scale $R$ is sufficiently large.  At late times $t \gtrsim \mathcal{O}(L^{2})$ the entropy relaxes towards an area-law value which we compute exactly. We propose a modified quasi-particle picture which accounts for all of these main features and agrees quantitatively well with our essentially exact numerical results. The quasiparticles are associated with the modes of a non-Hermitian effective Hamiltonian. At small wave-vector $k$, the quasiparticles transport entropy with a finite velocity, but have a lifetime scaling as $1/k^2$; the concurrence of these two conditions  leads directly to the observed $t^{1/2}$ growth. 
\end{abstract}

\maketitle

\begin{figure}[b]
    \centering
    \includegraphics[width = \columnwidth,height = 80pt]{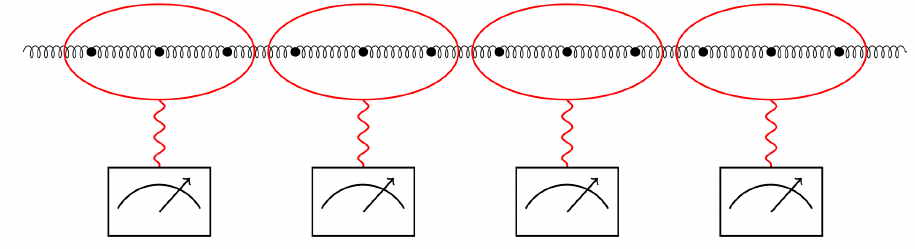}
    \caption{Cartoon picture of the setup for coarse-grained measurements of range $R = 3$. The red bubbles define blocks/unit cells in which the corresponding coarse-grained observable for the block has uniform support over all sites contained within. }
    \label{fig:Continuous weak monitoring diagram}
\end{figure}

\paragraph{Introduction.---} 
The dynamics of entanglement in many-body quantum systems is a topic of experimental \cite{Islam_2015,Google_2023,Brydges2019,Joshi2023} and theoretical interest, the study of which has led to new insights into how many-body systems come to (or fail to come to) equilibrium \cite{Bardarson_2012,Kim_2013,Ho_2017,Nahum_2017,Rakovszky_2019,Foligno_2023,Bertini_2022,doyon2023generalized}. More recently there has been a significant study of entanglement dynamics in measured systems \cite{Skinner_2019,Chan_2019,Buchhold_2021,ChargeSharpening}. In this context, measurements tend to reduce entanglement as opposed to unitary dynamics which increase it; the resulting competition can lead to a sharp transition in the late-time entanglement.
 
The `quasiparticle (QP) picture'  is a powerful heuristic which accounts for the growth of entanglement in integrable \emph{unitary} (not measured) systems; in this cartoon picture of many-body systems, entanglement grows due to the ballistic separation of EPR-correlated quasiparticle pairs. This picture initially appeared as a simplified description of exact CFT and free-fermion   calculations \cite{Calabrese_2005,Calabrese_2009}, but also provides a good quantitative description of more general integrable systems \cite{AlbaCalabrese2017}. The connection between entanglement growth and the quasiparticle picture has since been made  rigorous in specific cases \cite{Doyon2023}.   

It is natural to ask how measurements modify entanglement growth and the QP picture. Previous work \cite{Cao_2019,Turkeshi_2022} showed that in the case of free fermions with local density measurements, the growth of entanglement is described well by a modified QP picture which employs an ansatz: the action of measurements is to randomly `reset' the EPR pairs according to a Poisson process.  Here we study a non-interacting chain of harmonic oscillators subject to continuous weak monitoring of linear observables, starting from an unentangled Gaussian pure state. The measured observables are coarse-grained over $R$ adjacent lattice sites, so that as $R$  is increased individual measurements reveal less about the local correlations in the state. The resulting model can be solved semi-analytically by significantly extending the methods in \cite{Buchhold_2021,Minoguchi_2022}.  We propose a QP picture for entanglement growth in this model which differs considerably in its spirit and predictions from the `resetting' ansatz, and which we can justify semi-rigorously using the standard QP picture.  We verify this picture with essentially exact numerical and semi-analytical calculations.
 
For small measurement range ($R=1$), the model agrees with one already studied \cite{Buchhold_2021}, giving a brief period of entanglement growth that rapidly saturates to an area law. However, a prediction of our picture is that the growth of entanglement is extremely sensitive to the spatial extent of the measured observables: We explain why when the measurement range is large enough ($R>2$), entanglement grows sub-ballistically as $\sqrt{t}$ in our model. The underlying intuition is that sufficiently coarse measurements allow for the existence of long-lived ballistically propagating quasiparticles. These quasi-particles are indexed by a wavevector $k$, and at small  $k$ they decay at a rate $O(k^2)$. These observations imply that their net contribution to the entanglement at time $t$ goes as $\sim v t e^{-  \Gamma k^2 t}$. Summing over contributions from small $k$ gives rise to the claimed $\sqrt{t}$ growth.

In the following we introduce the effective Hamiltonian formalism to solve the weakly measured oscillator chain.  We then derive approximate expressions for the correlation functions of (small-$k$) slowly-relaxing modes that govern the long time dynamics. The dynamics of the slowly relaxing modes are used to motivate a modified quasiparticle description of entanglement spreading, which explains the diffusive growth of entanglement for sufficiently coarse measurements.

\paragraph{Effective Hamiltonian formalism.---} 
A pure state density matrix $\hat{\rho}(t) = |\psi(t)\rangle\langle \psi(t)|$ subject to Hamiltonian $\hat{H}$ dynamics,  and weak continuous measurement of observables $\{ \hat{O}_{b} \}$, evolves according to the quantum master equation \cite{Jacobs_2014}
\begin{equation}
\begin{split}
\label{eqn:Single replica stochastic master equation} 
d \hat{\rho}(t) &= dt \big( -i \big[\hat{H}, \hat{\rho}(t) \big] -\frac{\gamma}{2} \sum_{b} \big[ \hat{O}_{b}, \big[ \hat{O}_{b}, \hat{\rho}(t) \big] \big] \big) \\ &+ \sum_{b} dW_{b} \big\{ \hat{M}_{b,t}, \hat{\rho}(t) \big\}.
\end{split}
\end{equation}
Here $dW_{b}$ are independent Gaussian random variables with zero mean and variance $\gamma dt$, and $\gamma\geq 0$ is the measurement strength. The measurement operators are defined  
$\hat{M}_{b,t} \equiv \hat{O}_{b} - \text{Tr}(\hat{\rho}(t)\hat{O}_{b})$.
Eq.~\eqref{eqn:Single replica stochastic master equation} is therefore non-linear in $\hat{\rho}$, and not generally solvable. However, when the Hamiltonian is quadratic and the measured observables are linear, states that are initially Gaussian remain so under Eq.~\eqref{eqn:Single replica stochastic master equation}, and this allows for a partial solution of the dynamics.

Consider the harmonic chain in 1D  with continuous monitoring of linear combinations of the local oscillator positions $\hat{x}_{j}$ (Fig.~\ref{fig:Continuous weak monitoring diagram}); the oscillators have corresponding momenta $\hat{p}_{j}$. We are interested in how entanglement grows, starting from an unentangled Gaussian state. This question can be solved as follows. For the specific measured dynamics we consider, the entanglement entropy at a given time $t$ can be obtained entirely from the equal-time 2-point connected correlation functions between positions and momenta  (computed at the same time $t$). These in turn, evolve according a set of \textit{deterministic} but non-linear equations, which are numerically tractable, and can be approximated analytically in certain limits.

A straightforward but long calculation shows \cite{sm} that the time-dependent 2-point correlation functions are captured exactly if we evolve the density matrix  with a non-Hermitian Hamiltonian $\hat{\rho}(t) = e^{-i\hat{H}_{\text{eff}}t}\hat{\rho}(0)e^{i\hat{H}_{\text{eff}}^{\dagger}t}$, where
\begin{equation}
\label{eqn:Effective Hamiltonian explicit}
    \hat{H}_{\text{eff}} =\hat{H} - i\gamma \sum_{ij} M_{ij}\hat{x}_{i}\hat{x}_{j}.
\end{equation}
where $\hat{H}=\sum_{ij}\left(V_{ij}\hat{x}_{i}\hat{x}_{j} + \delta_{ij}\hat{p}_{i}\hat{p}_{j} \right)$ is the Hamiltonian of the unmonitored dynamics. For the harmonic chain we have $V_{ij} = -\nabla^{2}_{ij} + m^{2}\delta_{ij}$ and the matrix $M$ is a positive matrix that depends on the specific set of observables measured.   The coarse observables we monitor take the form $\hat{O}_{b}= \hat{x}_{bR+1} + \cdots + \hat{x}_{(b+1)R}$; their support does not overlap and they span the entire lattice of $L$ sites (see Fig.\ref{fig:Continuous weak monitoring diagram}).  In this case $M_{ij} = 1$ if sites $i,j$ lie inside the same block, and $M_{ij} = 0$ otherwise. It turns out that our results do not change qualitatively even if we allow these blocks defining the observables to overlap, so long as the underlying lattice translational symmetry is preserved. 

The effective Hamiltonian is translation invariant with size $R$ unit cell, hence can be block-diagonalised by first giving the local operators 2 indices ($\hat{x}_{i},\hat{p}_{i} \to \hat{x}_{b,j},\hat{p}_{b,j}$) denoting the block and sub-lattice, then Fourier transforming over the first index 
\begin{equation} 
\hat{H}_{\text{eff}} = \frac{1}{2}\sum_{k} \hat{\mathbf{\Psi}}^{\dagger}(k) H^{(k)} \hat{\mathbf{\Psi}}(k)
\end{equation}
($\hat{\mathbf{\Psi}}^{\dagger}H\hat{\mathbf{\Psi}} = \sum_{n,m} \hat{\Psi}^{*}_{n}H_{nm}\hat{\Psi}_{m}$) where $k \in (-\pi,\pi]$. We have defined spinors, $\hat{\mathbf{\Psi}}(k) = \begin{pmatrix}  \hat{\mathbf{a}}_{k}^{*} & \hat{\mathbf{a}}_{-k}\end{pmatrix}^{T}$ in terms of the vector of ladder operators $ \hat{\mathbf{a}}_{k} = \begin{pmatrix}\hat{a}_{k,0} & \cdots & \hat{a}_{k,R-1} \end{pmatrix}$, that are related to the Fourier components of the canonical position/momentum operators that appear in Eq.~\eqref{eqn:Effective Hamiltonian explicit} via $\hat{a}_{k,j} = (\hat{x}_{k,j} + i\hat{p}_{-k,j})/\sqrt{2}$. The canonical commutation relations are encoded by the matrix $C$, defined $[ \hat{\Psi}^{*}(k)_{n}, \hat{\Psi}(k)_{m}] = C_{nm}$, with $C= \mathrm{diag}(I,-I)$.

The matrices $H^{(k)}$ are in general non-Hermitian and cannot be made diagonal via a canonical transformation. However, we can construct a canonical transformation $\hat{\mathbf{\Psi}}(k) = W_{k}\hat{\mathbf{\Phi}}(k)$ that brings the effective Hamiltonian into form \cite{sm}
\begin{equation} \label{eq:eff_1}
\hat{H}_{\text{eff}} = \sum_{k} \hat{\mathbf{\Phi}}^{\dagger}(k) Z(k) \hat{\mathbf{\Phi}}(k),
\end{equation}
 where  $\hat{\mathbf{\Phi}}(k) = \begin{pmatrix} \hat{\mathbf{b}}_{k}^{*} & \hat{\mathbf{b}}_{-k} \end{pmatrix}^{T}$ and with $Z(k)$ upper triangular. The diagonal elements of $Z(k)$ are denoted $\begin{bmatrix} E_{0}(k), \cdots, E_{R-1}(k), E_{0}(-k), \cdots, E_{R-1}(-k)\end{bmatrix}$. We refer to the $E_j(k)$ as the complex bandstructure of the effective Hamiltonian.

Even though the matrix $Z(k)$ is non-diagonal, so contains more than just the bandstructure, we will show that the long time entanglement dynamics is in fact captured by keeping only the diagonal elements of $Z(k)$. Thus the properties of the bandstructure are vital for characterising the long time dynamics. Moreover, the $E_{j}(k)$ will later be interpreted as a complex dispersion relation for quasi-particles.

\paragraph{Bandstructure properties.---} 
The long-time dynamics of correlation functions is governed by the bandstructure $E_{j}(k)$ of the non-hermitian effective Hamiltonian, in particular the imaginary part sets the rate at which these correlation functions relax towards their steady-state values. 
For momenta $k = 0$ we have an exact formula for the quasiparticle energies for the different bands, $E_{j}(0) = 2\sqrt{4\sin^{2}\left(\frac{\pi j}{R}\right) + m^2 -i\gamma R \delta_{j,0}}$.

It is only the $j=0$ ``gapped band'' that has an imaginary part at $k=0$, which will correspond to an exponential decay (rate $\gamma$ ) of the corresponding $j=0$ correlations, and implies they only contribute to short-time transient dynamics. We focus then on the $j>0$ ``gapless'' bands, which have real eigenvalues at $k=0$, but which develop imaginary $O(k^2)$ components at nonzero $k$
\begin{equation} 
\label{eqn:Complex energy expansion}
E_{j}(k) = E_{j} + v_{j}k + \frac{1}{2}\delta_{j}k^{2} - i\Gamma_{j}k^{2} + \cdots 
\end{equation}
for real parameters $E_{j}$, $v_{j}$, $\delta_{j}$ and $\Gamma_{j}$ \cite{sm}. 
It is these small-$k$ modes in the gapless bands that govern the long-time entanglement growth. In the following section we will write down the equation of motion for the correlation functions of these modes, and derive an approximate solution in the long-time limit. In the process we will justify interpreting $E_{j}(k)$ as a quasiparticle dispersion. We will then semi-rigorously derive a modified quasiparticle picture in terms of this complex dispersion whereby the real part sets the velocity of the quasiparticles whilst the imaginary part governs the rate at which they decay.

\paragraph{Dynamics of correlation functions.---} 
Assume that the initial state has the same translation symmetry as the effective Hamiltonian (translation symmetry with unit-cell of size $R$). Then all 2-point connected correlations are encoded by
\begin{equation} 
\label{eqn: Replica antisymmetric correlation function}
\sigma(k,t)_{nm} = \frac{\text{tr}\big( \frac{1}{2}\big\{\hat{\Phi}(k)_{n},\hat{\Phi}^{*}(k)_{m}\big\} \hat{\rho}(t)\big)}{\text{tr}\big(\hat{\rho}(t)\big)},
\end{equation}
where $\hat{\rho}(t) = e^{-i\hat{H}_{\text{eff}}t}\hat{\rho}(0)e^{i\hat{H}_{\text{eff}}^{\dagger}t}$. $\sigma$ is a square matrix  of size $2R$, as each of the $R$ bands appears once for $+k$ and once for $-k$.
Taking the time derivative of Eq.~\eqref{eqn: Replica antisymmetric correlation function} yields a non-linear Riccati matrix evolution  equation
\begin{equation} 
\label{eqn:Riccati eqn}
\partial_{t} \sigma_s (k,t) = iCX  \sigma_s - i\sigma_s XC  - 2 \sigma_s Y\sigma_s -\{\sigma_s,Y\},
\end{equation}
where it is convenient to write the equation in terms of $\sigma_s \equiv \sigma-I/2$; this makes it clear that $\sigma=I/2$ is a steady-state.  In deriving this equation, we have used Wick's theorem to re-write 4-point correlators in terms of products of 2-point correlators. Moreover, the matrices $X,Y$ are Hermitian and defined through $Z(k)=X(k)- i Y(k)$. We have kept implicit the $k$ dependence of the r.h.s. of Eq.~\eqref{eqn:Riccati eqn} for brevity.

In the following we will deduce the entanglement dynamics by approximating Eq.~\eqref{eqn:Riccati eqn}, and we check our approximations with direct numerical integration (see \cite{sm}). To begin we analyse the Riccati equation at small $k$.  The terms in Eq.~\eqref{eqn:Riccati eqn} involving the matrix $X$ generate phase oscillations in the correlation functions, whilst terms involving $Y$ cause exponential relaxation towards the steady state  (i.e., $\sigma = I/2$ for each $k$). This exponential relaxation occurs on a short $\tau = \mathcal{O}(1)$ timescale for those elements of $\sigma_s$ involving at least one mode in the gapped band ($j = 0$), whilst all other correlation functions relax much slower on an $\mathcal{O}(1/k^2)$ timescale. This follows from our earlier observation that $E_{j}(k)$ has an imaginary gap at small $k$ precisely for $j=0$. 

Therefore, once $t \gtrsim \tau$, it is a good approximation to ignore correlations involving the gapped bands. Operationally, this means setting the $0$th/$R$th rows and columns of $\sigma_s$ to zero. This yields an approximate equation of motion for the remaining components of $\sigma_s$
\begin{equation}
\label{eqn:Ricatti equation expansion}
\begin{split}
    \partial_{t}\sigma_{s} &= i\left[ X_{0}C,\sigma_{s}\right] + ik\left[X_{1}C,\sigma_{s}\right] \\ 
    &+ ik^{2}\left( CX_{2}\sigma_{s} - \sigma_{s}X_{2}C\right) - k^{2} \left( \{\sigma_{s},Y_{2}\}  + 2\sigma_{s}Y_{2}\sigma_{s}\right).
\end{split}
\end{equation}
Note that we have performed a formal expansion of matrices $X,Y$ (e.g. $X(k) = X_{0} + kX_{1} + k^{2}X_{2} + \cdots $) keeping terms up to $\mathcal{O}(k^{2})$. 

The matrices $X_{0,1}$ are in fact diagonal and encode the $\mathcal{O}(1),\mathcal{O}(k)$ components of the bandstructure for the gapless bands, and generate phase oscillations in the correlation matrix at a rate of at least $\mathcal{O}(k)$. The remaining terms in Eq.~\eqref{eqn:Ricatti equation expansion} generate changes in $\sigma_{s}$ at a much slower $\mathcal{O}(k^2)$ rate. This separation of scales (similar to that underlying a Magnus expansion) suggests that we may approximate the matrices $X_{2},Y_{2}$ by their components that commute with $X_{0,1}$, which is their diagonal part. 

This diagonal approximation  should capture the time-averaged evolution of correlation functions at small $k$. Moreover, in the same limit, it is equivalent to the evolution generated by an effective Hamiltonian 
\begin{equation} 
\label{eqn:Effective Hamiltonian diagonal Fourier component}
\begin{split}
\hat{H}_{\text{eff}}  &= \sum_k \sum_{j = 1}^{R-1}\left[ E_{j}(k) \hat{b}_{k,j}^{\dagger}\hat{b}^{\phantom{\dagger}}_{k,j} + E_{j}(-k) \hat{b}_{-k,j}^{\dagger}\hat{b}^{\phantom{\dagger}}_{-k,j} \right] \\ 
&= \hat{H}_{U} - i\hat{H}_{D},
\end{split}
\end{equation}
which is formed by keeping only the diagonal part of $Z$ in Eq.~\eqref{eq:eff_1}, and ignoring the gapped bands.
At this point we recognise the operators $\hat{b}_{k,j}^{\dagger},\hat{b}^{\phantom{\dagger}}_{k,j}$ as the creation/annihilation operators for the `quasiparticles' of the effective theory at long times,  with $E_{j}(k)$ corresponding to their complex energy. 

$\hat{H}_{U,D}$ are both Hermitian, are defined through Eq.~\eqref{eqn:Effective Hamiltonian diagonal Fourier component}, and commute with one another. Under these simplified dynamics, the time evolution for the normalised density matrix can be considered the composition of a dissipative quantum map followed by a unitary evolution
\begin{equation}
\label{eqn:Dissipative/unitary quantum maps}
\begin{split}
 \hat{\rho}(0) &\rightarrow \hat{\rho}'= \frac{e^{-\hat{H}_{D} t}\hat{\rho}(0)e^{-\hat{H}_{D} t}}{\text{tr}\left(e^{-\hat{H}_{D} t}\hat{\rho}(0)e^{-\hat{H}_{D} t }\right)}
 \\
\hat{\rho}'&\rightarrow \hat{U}(t)\hat{\rho}' \hat{U}(t)^{\dagger},
\end{split}
\end{equation}
where $\hat{U}(t) = e^{-i\hat{H}_{U}t}$. In \cite{sm} we write down the corresponding transformation for the correlation matrix under these maps, which when combined together gives us the exact solution
\begin{equation} 
\label{eqn:Slow mode correlator evolution}
\sigma(k,t) = A -  B [e^{i\epsilon_{k}Ct}\sigma(k,0) e^{-i\epsilon_{k}Ct} + A ]^{-1} B.
\end{equation}
where $A\equiv  \coth(\Lambda_{k} t)/2$, and $B\equiv \text{cosech}(\Lambda_{k}t)/2 $. We define these objects and the correlation matrix in terms of the Hermitian, diagonal matrices $\epsilon_{k},\Lambda_{k}$ that encode the gapless quasiparticle energies, $\epsilon_{k} - i\Lambda_{k} = \text{diag}(E_{1}(k),\cdots,E_{R-1}(k),E_{1}(-k),\cdots,E_{R-1}(-k))$.

The long time limit yields exponential relaxation towards the steady state $\sigma \to \frac{1}{2}I$ as expected. The separation of the unitary and dissipative evolutions in   Eq.~\eqref{eqn:Dissipative/unitary quantum maps} leads to a modification of the quasiparticle picture that we now describe. 
\begin{figure}
    \centering
    \includegraphics[scale = 0.5]{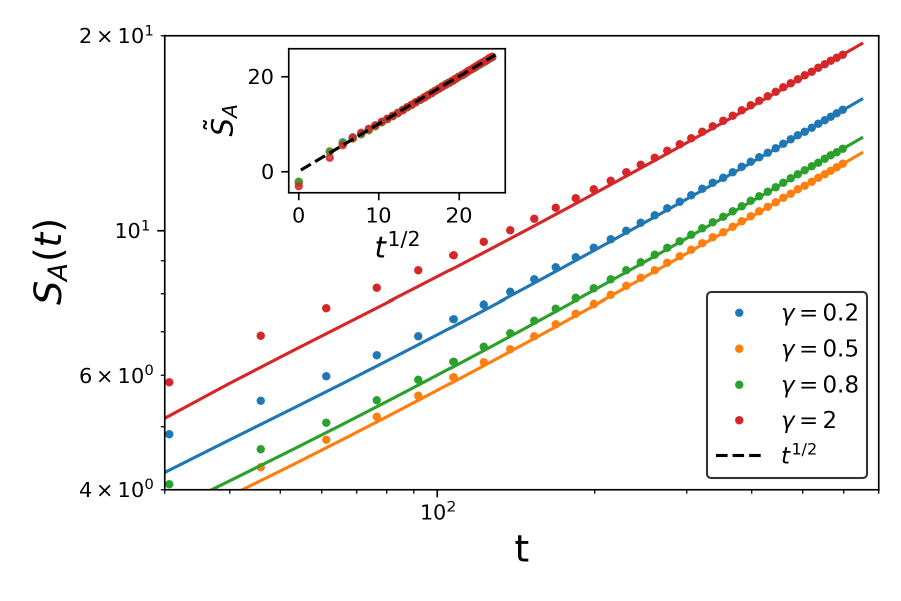}
    \caption{Half-chain entanglement entropy for a system of size $L=3000$ for measurements of range $R=4$, using periodic boundary conditions. The solid lines represents the quasiparticle picture predictions, while points are the result of directly integrating Eq.~\eqref{eqn:Riccati eqn}. The inset depicts a scaling collapse where we have fitted the $S_{A}(t)$ curves to the function $a + b\sqrt{t}$ and then plotted the shifted re-scaled entropy $\tilde{S}_{A} = (S_{A} - a)/b$.}
    \label{fig: Diffusive growth scaling collapse} 
\end{figure}

\paragraph{Quasiparticle picture.---} 
The standard quasiparticle picture for entanglement growth relies on the existence of stable and ballistically propagating quasiparticles associated with the densities $n_j(k)=\langle \hat{b}^{\dagger}_{k,j} \hat{b}^{\phantom{\dagger}}_{k,j}\rangle$, which are exactly conserved in closed systems. In our simple modification of this picture the quasiparticles have the \textit{complex} dispersion Eq.~\eqref{eqn:Complex energy expansion}; they  still propagate ballistically, but have a finite lifetime scaling as $O(1/k^{2})$ . 

To derive this picture,  recall that the small-$k$ dynamics can be split into dissipative/unitary parts Eq.~\eqref{eqn:Dissipative/unitary quantum maps}. The dissipative part of the evolution results in the decay of the quasiparticle densities
\begin{equation} 
\label{eqn:Exact density solution}
n_{j}(k,t) = \frac{n_{j}(k,0)e^{-4\Gamma_{j}k^{2} t}}{1 + n_{j}(k,0)(1 - e^{-4\Gamma_{j}k^{2} t})},
\end{equation}
which follows from Eq.~\eqref{eqn:Slow mode correlator evolution} for initial states $\sigma(k,t=0)$ having no inter-band correlations (a qualitatively accurate assumption discussed in \cite{sm}). $\Gamma_j $ derives from the leading imaginary contribution to $E_j(k)$ (Eq.~\eqref{eqn:Complex energy expansion}).

The only difference from the usual quasiparticle picture is that the initial densities are set by Eq.~\eqref{eqn:Exact density solution} (which note  depends on $t$). Under the unitary part of the dynamics, the quasiparticles spread ballistically and generate entanglement according to the usual quasiparticle picture: at each value of $k$, the quasiparticles form counterpropagating EPR pairs with velocities $\pm v_{j}(k) = \pm R \times \text{Re}(\partial_{k}E_{j}(k))$.

First consider the bipartite entanglement in an infinite system. The standard quasiparticle picture predicts that entropy grows as \cite{AlbaCalabrese2017}
\begin{equation} 
\label{eqn:Quasiparticle formula}
\begin{split}
S_{A}(t) = \; & \frac{1}{R}\sum_{j > 0}\int_{-\pi}^{\pi} \frac{dk}{2\pi}\; 2|v_{j}(k)| t \times \big[ - n_{j}(k,t) \log n_{j}(k,t) \\ &+\left(n_{j}(k,t) + 1\right) \log\left(n_{j}(k,t) + 1\right)\big].
\end{split}
\end{equation}
 The interpretation of Eq.~\eqref{eqn:Quasiparticle formula} forms our modified quasiparticle picture: the EPR pairs still transport entropy balistically at a rate set by their group velocity, however the quasiparticles forming these pairs have a finite lifetime set by the imaginary part of their complex energy (In our measured system, the $n_j(k,t)$ are no longer conserved but evolve according to Eq.~\eqref{eqn:Exact density solution}). Plugging these expressions into Eq.~\eqref{eqn:Quasiparticle formula} gives an integral dominated by $|k|\lesssim 1/\sqrt{\Gamma t}$ with a result
\begin{equation} \label{eq:sqrt}
S_{A}(t) \sim t^{1/2}\sum^{R-1}_{j > 0 } \frac{v_{j}}{\sqrt{\Gamma_{j}}} g(\nu_{j})
\end{equation}
Here $g\geq 0$ is a non-linear function of the conserved charge densities $\nu_{j} \equiv n_{j}(0,0)$.  Eq.~\eqref{eq:sqrt} predicts $\sqrt{t}$ growth of the entropy provided: i) $R>1$ and ii) $v_j g(\nu_j)\neq  0$ for some $j>0$. It turns out that  ii) is false when $R=2$, but generically true once $R>2$ \cite{sm}.  Eq.~\eqref{eq:sqrt} agrees well with the numerics obtained by direct integration of the Riccati equations (Fig.~\ref{fig: Diffusive growth scaling collapse}).

The above analysis assumed at various points that the quasiparticle bands are decoupled. This approximation is  quantitatively good for initial states with  translation symmetry (e.g., the state used for Fig.~\ref{fig: Diffusive growth scaling collapse}), and appears to also be qualitatively good (see \cite{sm,Bastianello_2018}). 

In a finite system of size $L$,  quasiparticle pairs traverse the system and reunite on a timescale $t_L=\frac{L}{2|v_{j}(k)|}$. This leads to oscillatory behavior in the bipartite entanglement \cite{Modak_2020,Santalla_2023}. To account for this, we must modify Eq.~\eqref{eqn:Quasiparticle formula}: replace the factor of $|v_{j}(k)| t$ in Eq.~\eqref{eqn:Quasiparticle formula}  by $f_{j}(k,L,t) \equiv \text{min}(2|v_{j}(k)|t,L - 2|v_{j}(k)|t)$, expressed in terms of the time modulo $t_L$ \cite{Chapman_2019} . The result is that the entanglement entropy grows diffusively up until a time $\mathcal{O}(L)$ before relaxing to the area-law steady state value as $S\sim L/\sqrt{t}$ on a timescale $\mathcal{O}(L^2)$.

\paragraph{Discussion.---}  
We examined a harmonic chain subjected to unitary dynamics and the weak monitoring of coarse-grained observables. We found a semi-analytical solution to the dynamics Eq.~\eqref{eqn:Slow mode correlator evolution}, by performing a novel long-wavelength analysis of the (Riccati) equations of motion for correlation functions. 

Using that, we show that when the measurements are sufficiently coarse, it allows for the existence of long-lived mode which lead to an unusual entanglement growth,  qualitatively much different than in the cases where the measured observables are finely-resolved, or in the absence of measurements. Specifically, we find that entanglement grows as $\sim\sqrt{t}$ when the measurements are smeared over $R > 2$ sites. We confirm these results numerically by directly integrating the Riccati equations. We explain our numerics with a novel quasiparticle picture: Diffusive entropy growth follows from the fact that (small $k$) quasiparticle modes transport entropy ballistically, but decay at a slow $\mathcal{O}(k^2)$ rate. 

The new quasiparticle picture predicts the asymptotic entanglement growth (Eq.~\eqref{eq:sqrt}) in terms of an $\mathcal{O}(R)$ number of coefficients $\left\{v_{j},\Gamma_{j},\nu_{j}\right\}$, which capture the pertinent features of the initial state and the  non-Hermitian quasiparticle dispersion.  Eq.~\eqref{eq:sqrt} often agrees quantitatively well with numerical simulations. 

Our specific methods are limited to the study of measured dynamics which preserve Gaussianity, however the quasiparticle picture can be applied to more general interacting integrable models \cite{AlbaCalabrese2017}. The effect of weak monitoring on integrable systems has to our knowledge not been studied; it would be interesting to investigate whether our modified quasiparticle picture can apply in this more general context.

\paragraph{Acknowledgements.---}  T.Y.  is supported by EPSRC studentship. C.K. is supported by a UKRI Future Leaders
Fellowship MR/T040947/1.

\bibliography{main}

\newpage
\onecolumngrid


\section*{Supplementary material}

# Supplementary material

Thomas Young,[1,2] Dimitri Gangardt,[1] and Curt von Keyserlingk[2]

[1]*School of Physics and Astronomy, University of Birmingham, Edgbaston, Birmingham, B15 2TT, UK*
[2]*Department of Physics, King's College London, Strand WC2R 2LS, UK*

(Dated: March 6, 2024)

## EFFECTIVE HAMILTONIAN DESCRIPTION

We are interested in calculating the evolution of the 2-point connected correlation functions

$$\tilde{\sigma}_{ij} = \text{tr}\left(\frac{1}{2}\{\hat{r}_i, \hat{r}_j\}\hat{\rho}\right) - \text{tr}(\hat{r}_i\hat{\rho})\text{tr}(\hat{r}_j\hat{\rho}) \tag{1}$$

where $\hat{\underline{r}} = (\hat{x}_1, \cdots, \hat{x}_L, \hat{p}_1, \cdots, \hat{p}_L)$. In [1] the introduction of 2 replica's of the density matrix was used to derive an effective Hamiltonian that was used to calculate the steady state entanglement in a system of free fermions with local density measurements. For completeness we include our modified version of their derivation which is applicable when the state is Gaussian.

We see that this correlation function is in fact non-linear in the density matrix $\hat{\rho}$, however it is in fact linear in the 2-replica density matrix $\hat{\rho}^{(R_2)} = \hat{\rho} \otimes \hat{\rho}$

$$\sigma_{ij} = \frac{1}{4}\text{tr}\left(\left\{\hat{r}_i^{(1)} - \hat{r}_i^{(2)}, \hat{r}_j^{(1)} - \hat{r}_j^{(2)}\right\}\hat{\rho}^{(R_2)}\right) \tag{2}$$

where we have given the canonical operators replica indices, i.e. $\hat{r}_i^{(1)} = \hat{r}_i \otimes \hat{I}, \hat{r}_i^{(2)} = \hat{I} \otimes \hat{r}_j$.

The 2-point *connected* correlation functions are functions only of the replica anti symmetric canonical operators $\hat{r}_i^{(-)}$ defined through

$$\hat{r}_i^{(\pm)} = \frac{\hat{r}_i^{(1)} \pm \hat{r}_i^{(2)}}{\sqrt{2}}. \tag{3}$$

For Gaussian states, the density matrix $\hat{\rho}$ is a Gaussian function of the canonical operators $\hat{r}_j$ which leads to the following

$$\hat{\rho}^{(R_2)} = \hat{\rho}^{(+)}\hat{\rho}^{(-)}. \tag{4}$$

We construct the following evolution equation for the 2-replica density matrix

$$d\hat{\rho}^{(R_2)} = d\hat{\rho} \otimes \hat{\rho} + \hat{\rho} \otimes d\hat{\rho} + d\hat{\rho} \otimes d\hat{\rho} \tag{5}$$

which leads to the following (remembering the Itô rule $dW_i dW_j = \gamma dt \delta_{ij}$)

$$d\hat{\rho}^{(R_2)} = dt\left(\mathcal{L}^{(1)} + \mathcal{L}^{(2)}\right)\hat{\rho}^{(R_2)} + \gamma dt \sum_b \left\{\hat{M}_{b,t}^{(1)}, \left\{\hat{M}_{b,t}^{(2)}, \hat{\rho}^{(R_2)}\right\}\right\} + \sum_b dW_b \left\{\hat{M}_{b,t}^{(1)} + \hat{M}_{b,t}^{(2)}, \hat{\rho}^{(R_2)}\right\} \tag{6}$$

where

$$\mathcal{L} \equiv -i\left[\hat{H}, \bullet\right] - \frac{\gamma}{2}\sum_b\left[\hat{O}_b, \left[\hat{O}_b, \bullet\right]\right]. \tag{7}$$

In terms of the replica symmetric/anti-symmetric operators this evolution equation takes the form

$$d(\hat{\rho}^{(+)}\hat{\rho}^{(-)}) = dt\left(\mathcal{L}^{(+)} + \mathcal{L}^{(-)}\right)\hat{\rho}^{(+)}\hat{\rho}^{(-)} + \frac{\gamma}{2}dt\sum_b\left\{\hat{M}_{b,t}^{(+)}, \left\{\hat{M}_{b,t}^{(+)}, \hat{\rho}^{(+)}\hat{\rho}^{(-)}\right\}\right\} - \frac{\gamma}{2}\sum_b\left\{\hat{O}_b^{(-)}, \left\{\hat{O}_b^{(-)}, \hat{\rho}^{(+)}\hat{\rho}^{(-)}\right\}\right\}$$
$$+ \sqrt{2}\sum_b dW_b \left\{\hat{M}_{b,t}^{(+)}, \hat{\rho}^{(+)}\hat{\rho}^{(-)}\right\} \tag{8}$$

where we have used the fact that replica symmetry implies $\text{tr}(\hat{O}_b^{(-)}\hat{\rho}^{(+)}\hat{\rho}^{(-)}) = 0$ meaning that $\hat{M}_{b,t}^{(-)} = \hat{O}_b^{(-)}$.

We notice that the product structure of the 2-replica density matrix is preserved by the evolution equation, hence we can consider the components $\hat{\rho}^{(\pm)}$ to evolve separately according to their own individual master equations. These separate master equations do not preserve the traces of the individual components, however the trace of their product is preserved $\text{tr}(\hat{\rho}^{(+)}\hat{\rho}^{(-)}) = \text{tr}(\hat{\rho}^{(+)})\text{tr}(\hat{\rho}^{(-)}) = 1$. This means that the 2-point *connected* correlation functions can be expressed in terms of the replica anti-symmetric part of the 2-replica density matrix only

$$\sigma_{ij} = \frac{\text{tr}\big(\frac{1}{2}\big\{\hat{r}_i^{(-)}, \hat{r}_j^{(-)}\big\} \hat{\rho}^{(-)}\big)}{\text{tr}(\hat{\rho}^{(-)})}. \tag{9}$$

Finally, we write down the evolution equation for this part for $\hat{\rho}^{(-)}$

$$\begin{aligned}\partial_t \hat{\rho}^{(-)} &= -i\left[\hat{H}^{(-)}, \hat{\rho}^{(-)}\right] - \frac{\gamma}{2}\sum_b \left[\hat{O}_b^{(-)}, \left[\hat{O}_b^{(-)}, \hat{\rho}^{(-)}\right]\right] - \frac{\gamma}{2}\sum_b \left\{\hat{O}_b^{(-)}, \left\{\hat{O}_b^{(-)}, \hat{\rho}^{(-)}\right\}\right\} \\ &= -i\left[\hat{H}^{(-)}, \hat{\rho}^{(-)}\right] - \gamma \sum_b \left\{\hat{O}_b^{(-)}\hat{O}_b^{(-)}, \hat{\rho}^{(-)}\right\}\end{aligned} \tag{10}$$

from which we obtain the effective non-hermitian Hamiltonian in the main text.

## TRANSFORMING THE EFFECTIVE HAMILTONIAN INTO UPPER TRIANGULAR FORM

We wish to construct the canonical transformation $W_k$ that makes the matrix $W_k^\dagger H^{(k)} W_k$ upper triangular. To start we consider the generalized eigenvalue equation for $H^{(k)}$

$$H\underline{e}_j = \lambda_j C \underline{e}_j \tag{11}$$

where we have dropped the index $k$ everywhere for brevity. We can construct the columns for the matrix $W_k$ using a generalized Gram-Schmidt procedure. These column vectors $\underline{x}_j$ must satisfy

$$\begin{aligned}\underline{x}_i^\dagger H \underline{x}_j &= 0 \quad i > j \\ \underline{x}_i^\dagger C \underline{x}_j &= C_{ij}\end{aligned} \tag{12}$$

where the second condition ensures that the basis transformation is canonical. The matrix $H$ has the following properties that we will make use of

$$\begin{aligned}\text{Re}\left(\underline{x}^\dagger H \underline{x}\right) &\geq 0 \\ \text{Im}\left(\underline{x}^\dagger H \underline{x}\right) &\leq 0.\end{aligned} \tag{13}$$

The matrices $H, C$ are square with dimension $2R$ and we choose the first $R$ vectors $\underline{e}_j$ to satisfy $\underline{e}_j^\dagger C \underline{e}_j = 1$ and the others to satisfy $\underline{e}_j^\dagger C \underline{e}_j = -1$.

The transformation matrix $W_k$ is parameterised by columns $\underline{x}_j$

$$W_k = \begin{pmatrix}\underline{x}_1 & \cdots & \underline{x}_{2R}\end{pmatrix} \tag{14}$$

The first column, $\underline{x}_1 \equiv \underline{e}_1$ is chosen to be one of the generalized eigenvectors. We require that the second column satisfies

$$\underline{x}_2^\dagger C \underline{x}_1 = 0 \tag{15}$$

which can be achieved for

$$\underline{x}_2 = \underline{e}_2 - \frac{\underline{x}_1^\dagger C \underline{e}_2}{\underline{x}_1^\dagger C \underline{x}_1} \underline{x}_1. \tag{16}$$

This vector is then normalised such that $\underline{x}_2^\dagger C \underline{x}_2 = C_{22}$. This procedure is repeated to generate all of the columns of the matrix $W_k$ via the general formula

$$\underline{x}_j = \underline{e}_j - \sum_{i<j} \frac{\underline{x}_i^\dagger C \underline{e}_j}{\underline{x}_i^\dagger C \underline{x}_i} \underline{x}_i \tag{17}$$

with appropriate normalisation of the new column vector $\underline{x}_j$ generated each step.

This procedure generates a set of column vectors for the transformation matrix that satisfy $\underline{x}_i^\dagger C \underline{x}_j = C_{ij}$. These vectors are related to the generalized eigenvectors of $H$ via

$$\underline{x}_i = \alpha_{ij} \underline{x}_j \tag{18}$$

with the matrix $\alpha$ upper triangular. This gives the following expression for the matrix elements of $H$ in the new basis

$$\begin{aligned}
\underline{x}_i^\dagger H \underline{x}_j &= \sum_{n=1}^{j} \sum_{m=1}^{n} \alpha_{nj} \alpha_{nm}^{-1} \lambda_n \underline{x}_i^\dagger C \underline{x}_m \\
&= \sum_{n=1}^{j} \sum_{m=1}^{n} \alpha_{nj} \alpha_{nm}^{-1} \lambda_n C_{im}
\end{aligned} \tag{19}$$

where we have used the fact that the matrix $\alpha^{-1}$ is also upper triangular. The RHS is zero for $i > j$ since the only terms in the sum that give non-zero contributions are $m = i$ however $m$ is bounded from above by $j$. This means that the matrix $W_k^\dagger H^{(k)} W_k$ is indeed upper triangular. It's diagonal elements are given by

$$\begin{aligned}
\underline{x}_i^\dagger H \underline{x}_i &= \alpha_{ii} (\alpha^{-1})_{ii} \lambda_i C_{ii} \\
&= \lambda_i C_{ii}
\end{aligned} \tag{20}$$

where we have used the fact that the eigenvalues of a triangular matrix are it's diagonal elements. The RHS of (20) is the quasiparticle energy.

## QUASIPARTICLE BANDSTRUCTURE

The bandstructure $E_j(k)$ is obtained simply from the generalized eigenvalues

$$H^{(k)} \underline{e}_j = \lambda_j(k) C \underline{e}_j. \tag{21}$$

The matrix $H^{(k)}$ has the block form

$$H^{(k)} = \begin{pmatrix} V^{(k)} + I - i\gamma M & V^{(k)} - I - i\gamma M \\ V^{(k)} - I - i\gamma M & V^{(k)} + I - i\gamma M \end{pmatrix} \tag{22}$$

in terms of the $(R \times R)$ matrices

$$V^{(k)} = \begin{pmatrix} 2 & -1 & 0 & \cdots & 0 & -e^{-ik} \\ -1 & 2 & -1 & \cdots & 0 & 0 \\ \vdots & \vdots & \vdots & \ddots & \vdots & \vdots \\ -e^{ik} & 0 & 0 & \cdots & -1 & 2 \end{pmatrix} + m^2 I \tag{23}$$

$$M = \begin{pmatrix} 1 & 1 & 1 & \cdots & 1 & 1 \\ 1 & 1 & 1 & \cdots & 1 & 1 \\ \vdots & \vdots & \vdots & \ddots & \vdots & \vdots \\ 1 & 1 & 1 & \cdots & 1 & 1 \end{pmatrix}. \tag{24}$$

Using this block structure, the generalized eigenvalue equation can be re-expressed

$$\left( V^{(k)} - i\gamma M \right) \tilde{\underline{x}}_j = \frac{\lambda_j(k)^2}{4} \tilde{\underline{x}}_j \tag{25}$$

from which we conclude that $\lambda_j(k)^2$ are the eigenvalues of the matrix $4(V^{(k)} - i\gamma M)$.

The bandstructure $E_j(k)$ is thus the eigenvalues of the matrices $2\sqrt{V^{(k)} - i\gamma M}$ where the square root returns the root with positive real part. For the special case of $k = 0$ the matrices $V^{(k)}$ and $M$ commute as they're both circulant, from which the formula in the main text can be simply obtained.

The expansion of the bandstructure around $k = 0$ can be obtained using perturbation theory to write a perturbative expansion for the eigenvalues of $V^{(k)} - i\gamma M$, in the momenta $k$. The unperturbed eigenvectors are fourier modes since $V^{(0)}, M$ are circulant. The perturbation can be expressed

$$\delta V = \begin{pmatrix} 0 & 0 & 0 & \cdots & 0 & 1-e^{-ik} \\ 0 & 0 & 0 & \cdots & 0 & 0 \\ \vdots & \vdots & \vdots & \ddots & \vdots & \vdots \\ 1-e^{ik} & 0 & 0 & \cdots & 0 & 0 \end{pmatrix} \tag{26}$$

which is in an inconvenient form since the matrix is not linear in the expansion parameter $k$. However, the parameters $v_j, \Gamma_j$ can be obtained by keeping only the part of the perturbation that is linear in $k$. In terms of this linear perturbation, the coefficient $v_j$ is obtained from the 1st order correction whilst $\Gamma_j$ is obtained from the imaginary part of the 2nd order correction. Applying perturbation theory for this linearised perturbation yields the following

$$v_j = \frac{4\sin\left(\frac{2\pi j}{R}\right)}{E_j(0)R} \tag{27}$$

$$\Gamma_j = \frac{8\sin^2\left(\frac{\pi j}{R}\right)}{E_j(0)R^2} \frac{\gamma R}{(\gamma R)^2 + \left(4\sin^2\left(\frac{\pi j}{R}\right)\right)^2} \tag{28}$$

with $E_j(0)$ being given in the main text.

Notice that if we set $R = 2$ then the group velocity for the only gapless band $(j = 1)$ is in fact zero which means that our modified quasiparticle picture would predict that there is no long time entanglement growth. This is consistent with our numerical simulation of the dynamics.

For $R > 2$ we find that the long time entanglement growth is diffusive. A natural question to ask is what is the timescale on which this diffusive behavior emerges. To answer this question we look at the quasiparticle picture formula for the entropy growth

$$S_A(t) = \frac{1}{R}\sum_{j>0}\int_{-\pi}^{\pi}\frac{dk}{2\pi} 2|v_j(k)|t \times \left[(n_j(k,t)+1)\log(n_j(k,t)+1) - n_j(k,t)\log n_j(k,t)\right] \tag{29}$$

and make a change of variables $q = k\sqrt{\Gamma_j t}$

$$\begin{aligned} S_A(t) &= t^{1/2}\sum_{j>0}\frac{2v_j}{R\sqrt{\Gamma_j}}\int_{-\pi\sqrt{\Gamma_j t}}^{\pi\sqrt{\Gamma_j t}}\frac{dq}{2\pi}\left[(n_j(q)+1)\log(n_j(q)+1) - n_j(q)\log n_j(q)\right] \\ &= t^{1/2}\sum_{j>0}\frac{v_j}{\sqrt{\Gamma_j}}\left(g(\nu_j) + \mathcal{O}(\sqrt{t}e^{-4\pi^2\Gamma_j t})\right) \end{aligned} \tag{30}$$

keeping only the leading order terms in $t$. We conclude that the diffusive behaviour emerges on a timescale $\mathcal{O}(1/\Gamma_j)$ which is in qualitative agreement with the figure in the main text. This timescale diverges in the limit of no measurements ($\gamma \to 0$) as we would expect since the entanglement growth under unitary dynamics is in general linear in time. This timescale also diverges in the limit $\gamma \to \infty$, the interpretation of which is not so obvious. In this limit the monitored observables are measured projectively every time step which leads to the quantum zeno effect. However, since only a fraction $1/R$ of the degrees of freedom of the system are measured we find that this quantum zeno effect is only partial, meaning that the remaining degrees of freedom evolve under effective unitary dynamics for which we would expect the entanglement growth generated to be linear in time. We find excellent agreement between the numerical simulations of strongly monitored dynamics and simulations for the effective unitary evolution of the un-monitored degrees of freedom, with the entanglement growing linearly in time.

## RICCATI EQUATION EXPANSION

In the main text we perform a formal expansion of the hermitian matrices $X(k), Y(k)$ that define the hermitian/anti-hermitian parts of our transformed effective Hamiltonian

$$W_k^\dagger H^{(k)} W_k \equiv Z(k) = X(k) - iY(k). \tag{31}$$

The canonical transformation, $W_k$, as well as the matrix $H^{(k)}$ have formal expansions

$$\begin{aligned} W_k &= W_0 + kW_1 + k^2 W_2 + \cdots \\ H^{(k)} &= H^{(0)} + kH^{(1)} + k^2 H^{(2)} + \cdots \end{aligned} \tag{32}$$

where importantly only the matrix $H^{(0)}$ in the second expansion is non-hermitian. Another important property that we will utilise is that the matrix $Z(k)$ is upper triangular at all orders in $k$.

We now define the 'fast/slow' subspaces which are the rows/columns of our matrices associated with the $j = 0 / j > 0$ band operators.

The zero-order term in the full expansion has the following 'block diagonal' form

$$W_0^\dagger H^{(0)} W_0 = Z_f \oplus Z_s \tag{33}$$

with respect to these subspaces, where $Z_s$ is hermitian (since the quasiparticle energies for the $j > 0$ bands have gapless imaginary parts) thus diagonal. This results in the matrix $X_0$ appearing in the main text being diagonal. This is because the columns of $W_0$ associated with the 'slow' subspace lie in the kernel of the non-hermitian part of $H^{(0)}$.

The first order term in the perturbative expansion takes the form

$$Z^{(1)} = W_0^\dagger H^{(1)} W_0 + W_1^\dagger H^{(0)} W_0 + W_0^\dagger H^{(0)} W_1 \tag{34}$$

where the first term is hermitian. The non-hermitian part of this perturbation is associated with the non-hermitian part of $H^{(0)}$. Thus in the block form of $Z^{(1)}$

$$Z^{(1)} = \begin{pmatrix} Z_f^{(1)} & Z_{fs}^{(1)} \\ Z_{sf}^{(1)} & Z_s^{(1)} \end{pmatrix} \tag{35}$$

the sub-matrix $Z_s^{(1)}$ is also in fact hermitian thus diagonal. This results in the matrix $X_1$ appearing in the main text being diagonal.

The fact that the matrices $Z_s^{(0)}, Z_s^{(1)}$ are hermitian also implies that in the expansion of $Z(k)_s$, non-hermiticity appears first at 2nd order hence there are no $Y_0, Y_1$ terms in our Riccati equation expansion.

## DERIVING THE RICCATI EQUATION

To derive the Riccati equation in the main text we start with the master equation for the density matrix

$$\partial_t \hat{\rho} = -i\left[\hat{\mathbf{\Phi}}_k^\dagger X \hat{\mathbf{\Phi}}_k, \hat{\rho}\right] - \left\{\hat{\mathbf{\Phi}}_k^\dagger Y \hat{\mathbf{\Phi}}_k, \hat{\rho}\right\}. \tag{36}$$

In terms of this density matrix, the correlation functions are defined

$$\sigma(k, t)_{nm} = \frac{\mathrm{tr}\left(\frac{1}{2}\{\hat{\Phi}(k)_n, \hat{\Phi}^*(k)_m\}\hat{\rho}(t)\right)}{\mathrm{tr}\left(\hat{\rho}(t)\right)}. \tag{37}$$

Taking the time derivative yields

$$\partial_t \sigma(k, t)_{nm} = \frac{\mathrm{tr}\left(\frac{1}{2}\{\hat{\Phi}(k)_n, \hat{\Phi}^*(k)_m\}\partial_t \hat{\rho}(t)\right)}{\mathrm{tr}\left(\hat{\rho}(t)\right)} - \frac{\mathrm{tr}\left(\frac{1}{2}\{\hat{\Phi}(k)_n, \hat{\Phi}^*(k)_m\}\hat{\rho}(t)\right)}{\mathrm{tr}\left(\hat{\rho}(t)\right)^2} \mathrm{tr}\left(\partial_t \hat{\rho}\right). \tag{38}$$

To begin to tackle the above expression we first derive a very important property of the anti-hermitian term in the effective Hamiltonian, using the defining property of the canonical transformation, $W_k^\dagger C W_k = C$, to obtain

$$\begin{aligned} \text{tr}(CY) &\propto \text{tr}(CW_k^\dagger \mathbf{1} W_k) \\ &= \text{tr}(W_k^{-1} C \mathbf{1} W_k) \\ &= \text{tr}(C\mathbf{1}) \\ &= 0 \end{aligned} \quad (39)$$

where $\mathbf{1}_{ij} = 1$. From this point onwards we will leave the $k, t$ arguments implicit for brevity.

We examine the first term in the equation of motion

$$\begin{aligned} \text{tr}\big(\tfrac{1}{2}\big\{\hat{\Phi}_n, \hat{\Phi}_m^*\big\} \partial_t \hat{\rho}\big) &= -i\, \text{tr}\big(\big[\tfrac{1}{2}\big\{\hat{\Phi}_n, \hat{\Phi}_m^*\big\}, \hat{\mathbf{\Phi}}^\dagger X \hat{\mathbf{\Phi}}\big]\hat{\rho}\big) \\ &\quad - \text{tr}\big(\big\{\tfrac{1}{2}\big\{\hat{\Phi}_n, \hat{\Phi}_m^*\big\}, \hat{\mathbf{\Phi}}^\dagger Y \hat{\mathbf{\Phi}}\big\}\hat{\rho}\big) \end{aligned} \quad (40)$$

and notice that the 2nd part will require the use of Wick's theorem, for which we first need to put the quartic term in a more symmetrical form. We start by using the identity

$$\begin{aligned} \big\{\hat{\Phi}_n \hat{\Phi}_m^*, \hat{\Phi}_a^* \hat{\Phi}_b\big\} &= \tfrac{1}{2}\big(\big[\big[\hat{\Phi}_n \hat{\Phi}_m^*, \hat{\Phi}_a^*\big], \hat{\Phi}_b\big]\big) + \tfrac{1}{2}\big(\big\{\big\{\hat{\Phi}_n \hat{\Phi}_m^*, \hat{\Phi}_a^*\big\}, \hat{\Phi}_b\big\}\big) + C_{ab} \hat{\Phi}_n \hat{\Phi}_m^* \\ &= \tfrac{1}{2}\big(\big\{\big\{\hat{\Phi}_n \hat{\Phi}_m^*, \hat{\Phi}_a^*\big\}, \hat{\Phi}_b\big\}\big) - \tfrac{1}{2} C_{na} C_{mb} + C_{ab} \hat{\Phi}_n \hat{\Phi}_m^* \end{aligned} \quad (41)$$

alongside the identity proved above ($\text{tr}(CY) = 0$) to obtain

$$\text{tr}\big(\big\{\tfrac{1}{2}\big\{\hat{\Phi}_n, \hat{\Phi}_m^*\big\}, \hat{\mathbf{\Phi}}^\dagger Y \hat{\mathbf{\Phi}}\big\}\hat{\rho}\big) = -\tfrac{1}{2} C_{na} Y_{ab} C_{mb} \text{tr}(\hat{\rho}) + \tfrac{1}{2} \text{tr}\big(\big\{\big\{\tfrac{1}{2}\big\{\hat{\Phi}_n, \hat{\Phi}_m^*\big\}, \hat{\Phi}_a^*\big\}, \hat{\Phi}_b\big\}\hat{\rho}\big) Y_{ab} \quad (42)$$

which is in a nice form since the operators appear in a more symmetric way so Wick's theorem can be more easily utilised

$$\frac{1}{4} \frac{\text{tr}\big(\big\{\big\{\big\{\hat{\Phi}_n, \hat{\Phi}_m^*\big\}, \hat{\Phi}_a^*\big\}, \hat{\Phi}_b\big\}\hat{\rho}\big)}{\text{tr}(\hat{\rho})} Y_{ab} = 2\big(\sigma_{na}\sigma_{bm} + \sigma_{nm}(\sigma_{ba} + \tfrac{1}{2} C_{ab})\big) Y_{ab} \quad (43)$$

where we have used the fact that

$$\text{tr}\big(\hat{\Phi}_a \hat{\Phi}_b \hat{\rho}\big) = 0. \quad (44)$$

We next focus on the 2nd term in Eq. (38)

$$\begin{aligned} \frac{\text{tr}\big(\tfrac{1}{2}\{\hat{\Phi}_n, \hat{\Phi}_m^*\}\hat{\rho}\big)}{\text{tr}(\hat{\rho})^2} \text{tr}(\partial_t \hat{\rho}) &= -\sigma_{nm} \frac{\text{tr}\big(\{\hat{\Phi}_a^* \hat{\Phi}_b, \hat{\rho}\}\big)}{\text{tr}(\hat{\rho})} Y_{ab} \\ &= -2\sigma_{nm}\big(\sigma_{ba} + \tfrac{1}{2} C_{ab}\big) Y_{ab} \end{aligned} \quad (45)$$

which we see exactly cancels the 2nd term in Eq. (43) which it should since the correlation matrix is invariant upon changes in the norm of the density matrix.

Finally we look at the term associated with the hermitian part of the effective Hamiltonian (first term in Eq. (40))

$$\text{tr}\big(\big[\tfrac{1}{2}\big\{\hat{\Phi}_n, \hat{\Phi}_m^*\big\}, \hat{\mathbf{\Phi}}^\dagger X \hat{\mathbf{\Phi}}\big]\hat{\rho}\big) = -C_{an} X_{ab} \text{tr}\big(\tfrac{1}{2}\{\hat{\Phi}_b, \hat{\Phi}_m^*\}\hat{\rho}\big) + C_{mb} X_{ab} \text{tr}\big(\tfrac{1}{2}\{\hat{\Phi}_n, \hat{\Phi}_a^*\}\hat{\rho}\big). \quad (46)$$

Combining all of the results together yields

$$\partial_t \sigma(k,t)_{nm} = iC_{na} X_{ab} \sigma_{bm} - i\sigma_{na} X_{ab} C_{bm} - 2\sigma_{na} Y_{ab} \sigma_{bm} + \tfrac{1}{2} C_{na} Y_{ab} C_{bm} \quad (47)$$

which is the Riccati equation in the main text.

# RICCATI EQUATION SOLUTION

We wish to calculate the time dependent correlation matrix $\sigma_s(k,t)$ that corresponds to the transformation of the density matrix under the quantum map

$$\hat{\rho}_k(0) \to \frac{e^{-\hat{H}_D(k)t}\hat{\rho}_k(0)e^{-\hat{H}_D(k)t}}{\text{tr}\left(e^{-\hat{H}_D(k)t}\hat{\rho}_k e^{-\hat{H}_D(k)t}\right)} \quad (48)$$
$$\to \hat{U}(k,t)\hat{\rho}_k(0)\hat{U}(k,t)^\dagger.$$

To do so we first introduce the characteristic function representation of a quantum operator. For an $N$ mode bosonic quantum system we have a set of creation/annihilation operators defined by the commutation relations

$$\begin{aligned}\left[\hat{a}_i,\hat{a}_j^\dagger\right] &= \delta_{ij} \\ [\hat{a}_i,\hat{a}_j] &= 0.\end{aligned} \quad (49)$$

Any operator, $\hat{O}$ on the Hilbert space omits the following fourier esque decomposition

$$\hat{O} = \int d^{2N}z\, \chi(-z)\hat{D}(z) \quad (50)$$

in terms of the displacement operator $\hat{D}(z) = \exp\left(z_i\hat{a}_i^\dagger - \bar{z}_i\hat{a}_i\right)$ and the characteristic function $\chi(z) = \text{Tr}\left(\hat{O}\hat{D}(z)\right)$. The special feature of Gaussian operators is that their characteristic function is also Gaussian

$$\chi(z) \propto \exp\left(-\frac{1}{2}\left(\bar{z},z\right)C\sigma C\begin{pmatrix}z\\\bar{z}\end{pmatrix}\right) \quad (51)$$

where $\sigma$ is the symmetrically ordered correlation matrix for operators $\hat{a}_i, \hat{a}_i^\dagger$, and $C = \begin{pmatrix}I & 0\\0 & -I\end{pmatrix}$.

The goal is to calculate the transformation of the correlation matrix under the quantum map

$$\hat{\rho} \to \frac{\hat{M}\hat{\rho}\hat{M}}{\text{tr}\left(\hat{M}\hat{\rho}\hat{M}\right)} \quad (52)$$

with

$$\hat{M} \propto \exp\left(-\sum_j \lambda_j \hat{a}_j^\dagger \hat{a}_j\right) \quad (53)$$

for $\lambda_j \geq 0$. The transformation represents the dissipative part of the evolution in the main text. This operator is Gaussian hence has a Gaussian characteristic function of the form (51) with

$$\sigma = \frac{1}{2}\coth\left(\frac{\lambda}{2}\right) \equiv \sigma_M \quad (54)$$

where $\lambda = \text{diag}\left(\lambda_1,\cdots,\lambda_N\right)$.

The normalisation of the density matrix is not important for the calculation of the resulting correlation matrix. The transformed density matrix has the characteristic function representation

$$\hat{M}\hat{\rho}\hat{M} = \int d^{2N}z_1 d^{2N}z_2 d^{2N}z_3\, \chi_M(z_1)\chi_\rho(z_2)\chi_M(z_3)\hat{D}(z_1)\hat{D}(z_2)\hat{D}(z_3). \quad (55)$$

The transformed correlator can be calculated as follows; the operator exponentials are combined to give a single operator exponential, using the Baker-Campbell Hausdorff formula. The variables $\{z_1, z_2, z_3\}$ are transformed to a new basis in which only one of them appears in the operator exponential. The other 2 combinations are now 'dummy variables' since they do not appear in the operator exponential. These dummy variables can be removed by integrating over them, which can be done exactly since the integrand is Gaussian in these variables. Finally, we are left with an

expression of the form (50) where we can read off the transformed correlation matrix since we know that the operator $\hat{M}\hat{\rho}\hat{M}$ is Gaussian.

The displacement operators can be combined using

$$
\begin{aligned}
\hat{D}(z_1)\hat{D}(z_2)\hat{D}(z_3) &= \hat{D}(z_1 + z_2 + z_3) \\
&\times \exp\left(\frac{1}{2}\begin{pmatrix}\bar{z}_2 & z_2\end{pmatrix} C \begin{pmatrix} z_3 - z_1 \\ \bar{z}_3 - \bar{z}_1 \end{pmatrix} + \frac{1}{2}\begin{pmatrix}\bar{z}_1 & z_1\end{pmatrix} C \begin{pmatrix} z_3 \\ \bar{z}_3 \end{pmatrix}\right)
\end{aligned}
\tag{56}
$$

whilst the Gaussian integration over the dummy variables can be done using the formulae

$$
\begin{aligned}
\int d^{2N}z \ \exp\left(-\begin{pmatrix}\bar{z} & z\end{pmatrix} A \begin{pmatrix} z \\ \bar{z} \end{pmatrix} + \begin{pmatrix}\bar{z} & z\end{pmatrix}\begin{pmatrix} w \\ \bar{w} \end{pmatrix} \pm \begin{pmatrix}\bar{w} & w\end{pmatrix}\begin{pmatrix} z \\ \bar{z} \end{pmatrix}\right) \\
\propto \exp\left(\pm \begin{pmatrix}\bar{w} & w\end{pmatrix} A^{-1} \begin{pmatrix} w \\ \bar{w} \end{pmatrix}\right).
\end{aligned}
\tag{57}
$$

Following a straightforward but tedious calculation, the correlation matrix of the transformed density matrix can be read off, yielding the result in the main text once we set $\lambda = \Lambda_k t$ and account for the unitary part of the dynamics which is a known result $\sigma \to e^{i\epsilon_k C t} \sigma e^{-i\epsilon_k C t}$.

### LONG TIME AREA LAW STATE

As mentioned in the main text we find the infinite time steady state values for the correlation matrix

$$
\lim_{t\to\infty} \sigma(k,t) = \frac{1}{2} I \tag{58}
$$

which holds for all $k$ except $k = 0, \pi$ as there exists conserved quantities at these momenta that are decoupled from the dissipation so do not exhibit relaxation dynamics. We consider entanglement growth starting from initial states in which the conserved charge densities are finite in the thermodynamics limit.

Even though the effective Hamiltonian $\hat{H}_{\text{eff}}$ is not diagonalizable, the upper triangular form in the quasiparticle basis means that the quasiparticle vacuum state $|\psi_D\rangle$ satisfies

$$
: \hat{H}_{\text{eff}} : |\psi_D\rangle = 0 \tag{59}
$$

where $: :$ denotes normal ordering.

We consider first the entanglement entropy of this vacuum state who's correlation matrix is $\sigma(k) = \frac{1}{2} I \ \forall \ k$. From the fact that the canonical transformations $W_k$ are smooth in the momenta $k$ we can infer that the real space correlation functions decay exponentially hence the state $|\psi_D\rangle$ has area-law entanglement.

As mentioned in the main text, the presence of conserved charges that are completely decoupled from the dissipation at long times means that the long time state of the system under the effective Hamiltonian dynamics is in fact not the quasiparticle vacuum state

$$
\lim_{t\to\infty} \frac{e^{-i\hat{H}_{\text{eff}} t}|\psi(0)\rangle}{\sqrt{\langle\psi(0)|e^{i\hat{H}_{\text{eff}}^\dagger t} e^{-i\hat{H}_{\text{eff}} t}|\psi(0)\rangle}} \neq |\psi_D\rangle \tag{60}
$$

but has an $\mathcal{O}(1)$ density of quasiparticle excitations associated with the finite conserved charge densities. A consequence of these excitations that have infinite lifetime is that strictly speaking the system does not reach a steady state $|\psi_D\rangle$ (unless the initial sate has a 0 expectation value for these conserved charges). As long as these conserved charge densities are finite in the thermodynamic ($L \to \infty$) limit the long time entanglement scaling will still be area-law.

The presence of the finite conserved charge densities means that the momentum space correlation functions are no longer smooth in $k$ meaning that our previous explanation for the area-law entanglement scaling for the vacuum state $|\psi_D\rangle$ no longer applies. For example we consider the expectation value of the quasiparticle charge density

$$
\lim_{t\to\infty} \frac{\langle\psi(t)|\hat{b}_{k,j}^\dagger \hat{b}_{k,j}|\psi(t)\rangle}{\langle\psi(t)|\psi(t)\rangle} = \nu_j \delta(k) \tag{61}
$$

where $\delta(k) = 1$ for $k = 0$ and $\delta(k) = 0$ otherwise, with $\nu_j$ being the expectation value of the conserved charge in band $j$. We wish to quantify the effect that these discontinuities have on the entanglement entropy for which it is convenient to define the following correlation matrices

$$\sigma_{ij}^{(AB)} = \frac{1}{2}\langle\{\hat{A}_i, \hat{B}_j\}\rangle - \langle\hat{A}_i\rangle\langle\hat{B}_j\rangle \tag{62}$$

where $\hat{A}, \hat{B} \in [\hat{x}, \hat{p}]$ and $i, j$ denotes the lattice sites. Since the state has lattice translational symmetry with a unit cell of size $R$, these matrices are block circulant

$$\sigma^{(AB)} = \begin{pmatrix} \sigma_0^{(AB)} & \sigma_1^{(AB)} & \sigma_2^{(AB)} & \cdots & \sigma_2^{(AB)} & \sigma_1^{(AB)} \\ \sigma_1^{(AB)} & \sigma_0^{(AB)} & \sigma_1^{(AB)} & \cdots & \sigma_3^{(AB)} & \sigma_2^{(AB)} \\ \vdots & \vdots & \vdots & \ddots & \vdots & \vdots \\ \sigma_1^{(AB)} & \sigma_2^{(AB)} & \sigma_3^{(AB)} & \cdots & \sigma_1^{(AB)} & \sigma_0^{(AB)} \end{pmatrix} \tag{63}$$

The effects of the finite conserved charge densities is additive at the level of the correlation functions, hence we can write

$$\sigma^{(AB)} = \sigma_{\text{VAC}}^{(AB)} + \delta\sigma^{(AB)} \tag{64}$$

where $\sigma_{\text{VAC}}^{(AB)}$ is the correlation matrix for the area-law vacuum state $|\psi_D\rangle$. Since the vacuum state has exponentially decaying correlation functions in real space we have

$$\left(\sigma_{\text{VAC}}^{(AB)}\right)_x \approx 0 \quad \{x \gg \mathcal{O}(\zeta)\} \tag{65}$$

for some correlation length $\zeta$. To calculate the entanglement entropy between a region containing $N_A$ unit cells and it's complement in a systems comprised of $N$ unit cells we do the following; define the reduced correlation matrices that containing only correlations between sites restricted to the region $A$

$$\sigma^{(AB)}(N_A) = \begin{pmatrix} \sigma_0^{(AB)} & \sigma_1^{(AB)} & \sigma_2^{(AB)} & \cdots & \sigma_{N_A-2}^{(AB)} & \sigma_{N_A-1}^{(AB)} \\ \sigma_1^{(AB)} & \sigma_0^{(AB)} & \sigma_1^{(AB)} & \cdots & \sigma_{N_A-3}^{(AB)} & \sigma_{N_A-2}^{(AB)} \\ \vdots & \vdots & \vdots & \ddots & \vdots & \vdots \\ \sigma_{N_A-1}^{(AB)} & \sigma_{N_A-2}^{(AB)} & \sigma_{N_A-3}^{(AB)} & \cdots & \sigma_1^{(AB)} & \sigma_0^{(AB)} \end{pmatrix} \tag{66}$$

calculate the eigenvalues of the following matrix

$$\sigma^{(xx)}(N_A)\sigma^{(pp)}(N_A) - \sigma^{(xp)}(N_A)\sigma^{(px)}(N_A) \equiv \Lambda(N_A) \tag{67}$$

which we denote as $\{\lambda_1^2, \cdots, \lambda_{N_A+1}^2\}$.

The von-neumann entanglement entropy is calculated from the eigenvalues

$$S(N_A) = \sum_{j=1}^{N_A}\left[\left(|\lambda_j| + \frac{1}{2}\right)\log\left(|\lambda_j| + \frac{1}{2}\right) - \left(|\lambda_j| - \frac{1}{2}\right)\log\left(|\lambda_j| - \frac{1}{2}\right)\right]. \tag{68}$$

From the fact that the vacuum state has exponentially decaying correlation functions in real space, we would expect contributions to the entanglement entropy from sites that are within an $\mathcal{O}(\zeta)$ distance from the boundary of region $A$. For $N_A \gg \zeta$ we would expect the eigenvalues $\lambda_j$ calculated from the correlation functions between sites in the bulk to be equal to $\frac{1}{2}$ since these sites do not contribute to the entanglement entropy. This suggests the following approximation for the matrix from which these eigenvalues are calculated

$$\sigma_{\text{VAC}}^{(xx)}(N_A)\sigma_{\text{VAC}}^{(pp)}(N_A) - \sigma_{\text{VAC}}^{(xp)}(N_A)\sigma_{\text{VAC}}^{(px)}(N_A) \approx \Lambda_{\text{boundary}} \oplus \frac{1}{4}I_{\text{bulk}}(N_A) \tag{69}$$

where $\dim(\Lambda_{\text{boundary}}) = \mathcal{O}(\zeta)$, $\dim(I_{\text{bulk}}) = R(N_A - \mathcal{O}(\zeta))$. As a function of $N_A$ the subsystem entanglement entropy plateaus for $N_A \gg \mathcal{O}(\zeta)$ since contributions to the entanglement entropy only comes from sites near the boundary, which does not grow with $N_A$ in this limit.

We consider now the effect that the finite conserved charge density has on the subsystem entanglement scaling. Firstly we consider the number of unit cells comprising the system, $N$, to be odd meaning that there are only conserved charge densities at $k = 0$, each associated with one of the $R-1$ gapless bands. This means that the perturbation to the correlation matrices $\delta\sigma^{(AB)}$ are low rank, in particular

$$\delta\sigma^{(AB)} = \frac{1}{N} \begin{pmatrix} \delta\tilde{\sigma}_0^{(AB)} & \delta\tilde{\sigma}_0^{(AB)} & \cdots & \delta\tilde{\sigma}_0^{(AB)} \\ \delta\tilde{\sigma}_0^{(AB)} & \delta\tilde{\sigma}_0^{(AB)} & \cdots & \delta\tilde{\sigma}_0^{(AB)} \\ \vdots & \vdots & \ddots & \vdots \\ \delta\tilde{\sigma}_0^{(AB)} & \delta\tilde{\sigma}_0^{(AB)} & \cdots & \delta\tilde{\sigma}_0^{(AB)} \end{pmatrix} = \frac{1}{N} \delta\tilde{\sigma}_0^{(AB)} \otimes \mathbb{1}_N \quad (70)$$

where $\mathbb{1}_{ij} = 1$ for $1 \leq i,j \leq N$.

Using the fact that the correlations of $|\psi_D\rangle$ are exponentially decaying in real space we make the following approximation

$$\sum_{j=-\mathcal{O}(\zeta)}^{\mathcal{O}(\zeta)} \sigma_j^{(AB)} \approx \sum_{j=-\frac{N-1}{2}}^{\frac{N-1}{2}} \sigma_j^{(AB)} \equiv \tilde{\sigma}_0^{(AB)} \quad (71)$$

from which we conclude that the bulk part of the perturbed matrix $\Lambda(N_A)$ takes the form

$$\Lambda_{\text{bulk}}(N_A) = \frac{1}{4} I_{\text{bulk}}(N_A) + \frac{1}{N} A_0 \otimes \mathbb{1}_{N_{\text{bulk}}} + \frac{N_{\text{bulk}}}{N^2} B_0 \otimes \mathbb{1}_{N_{\text{bulk}}} \quad (72)$$

where $\dim(\mathbb{1}_{\text{bulk}}) = N_A - \mathcal{O}(\zeta)$ and the matrices $A_0, B_0$ are quadratic in $\delta\tilde{\sigma}_0^{(AB)}, \tilde{\sigma}_0^{(AB)}$.

From the fact that the state is pure before and after the perturbation, it can be shown that $A_0 + B_0 = 0$ hence we simplify

$$\Lambda_{\text{bulk}}(N_A) = \frac{1}{4} I_{\text{bulk}}(N_A) + \frac{1}{N}\left(1 - \frac{N_{\text{bulk}}}{N}\right) A_0 \otimes \mathbb{1}_{N_{\text{bulk}}}. \quad (73)$$

From this expression we see that most of the eigenvalues in the bulk part of $\Lambda(N_A)$ are unchanged, however a few of the bulk eigenvalues do in fact change, and they correspond to the eigenvalues of the matrix

$$\sqrt{\frac{1}{4}I + \frac{N_{\text{bulk}}}{N}\left(1 - \frac{N_{\text{bulk}}}{N}\right)A_0} \approx \sqrt{\frac{1}{4}I + \frac{N_A}{N}\left(1 - \frac{N_A}{N}\right)A_0} \quad (74)$$

from which we conclude that the effect of the perturbation is to modify the entanglement entropy profile

$$S(N_A) = S_{\text{VAC}}(N_A) + g\left(\frac{N_A}{N}\left(1 - \frac{N_A}{N}\right)\right) + \mathcal{O}(1/N) \quad (75)$$

for some function $g(x)$. The $\mathcal{O}(1/N)$ term is a correction to the boundary contribution to the entropy (The boundary part of the matrix $\Lambda$ is perturbed by a matrix with eigenvalues $\mathcal{O}(1/N)$). We would therefore expect that the presence of the finite conserved charge densities in the long time state of the system results in the subsystem entanglement profile having a parabolic like profile rather than a plateau for $N_A \gg \mathcal{O}(\zeta)$. The entanglement scaling is still area-law since the subsystem size appears in the entanglement entropy as the ratio $N_A/N$.

This parabolic profile is indeed what we see in the numerical calculations of the long time subsystem entanglement profile (Fig. 1)

In general determining the function $g$ is difficult, however in the limit in which the perturbations (conserved charge densities $\nu_j$) are large it can be shown that the entropy scales as

$$S(N_A) \approx \frac{\text{Rank}(A_0)}{2} \log \frac{N_A}{N}\left(1 - \frac{N_A}{N}\right) + \frac{1}{2} \log \text{pdet}(A_0) \quad (76)$$

where $\text{pdet}(A_0)$ is the pseudo-determinant (product of non-zero eigenvalues) of $A_0$. Here $\text{Rank}(A_0)$ is the number of conserved charges, which is equal to $R-1$, the number of gapless bands. We find excellent agreement between the analytical formula and the numerics in the limit where the perturbation is large (Fig. 2). Strictly speaking by large we mean that the increase in entanglement associated with these conserved charges is large compared to the vacuum state entanglement.

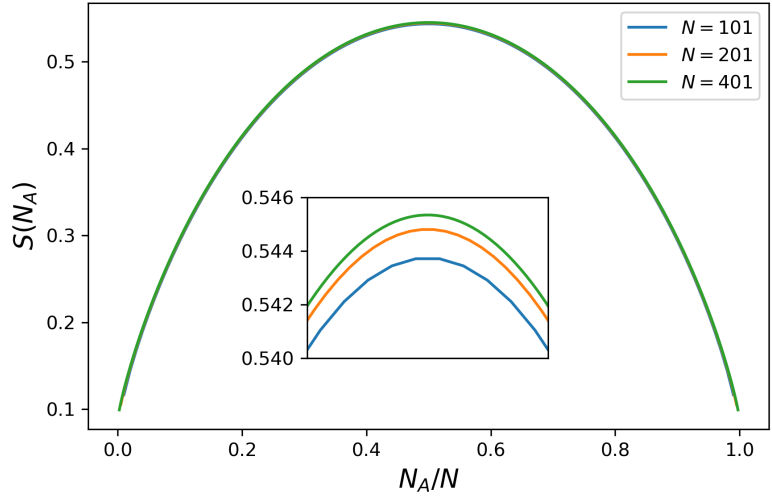

FIG. 1. Subsystem entanglement profile for the infinite time state for measurements of range 4 with $\gamma = 0.5$. The inset shows that the curve for different system sizes do not line up exactly. The finite size correction is observed numerically to be $\mathcal{O}(1/N)$ which is consistent with our prediction for the finite size effect.

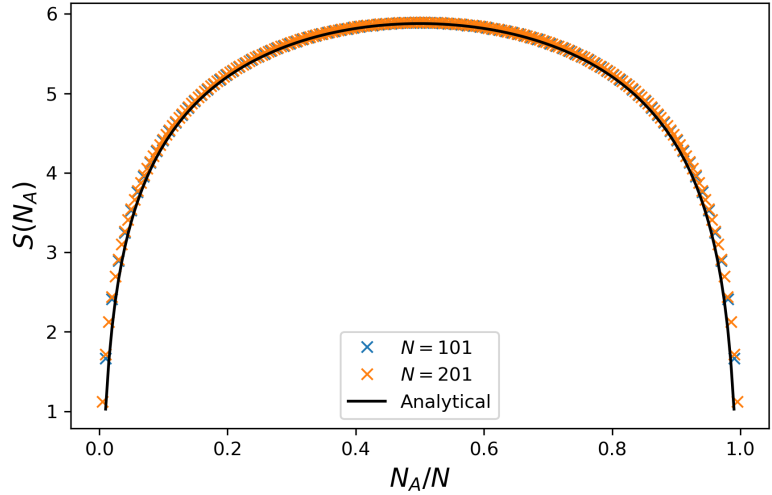

FIG. 2. Subsystem entanglement profile for the infinite time state in the limit where the conserved charge densities are large. We see excellent agreement with the analytical formula once the subsystem is large enough for the bulk/boundary separation to apply.

### DETAILS ON NUMERICAL SIMULATIONS

As discussed in the main text, the Gaussian property of the state can be used to express the dynamics of interest in terms of a set of Riccati equations for the correlation matrix, which are non-linear first order ODE's. The solution to these ODE's can be calculated using various numerical techniques, however one of the primary issues with such methods is that the correlation matrix must remain 'normalized' which complicates these methods.

The normalization condition arises since the quantum state is pure meaning that it has zero entanglement entropy. For a state with correlation matrix

$$\sigma_{ij} = \frac{1}{2}\langle\{\hat{r}_i, \hat{r}_j\}\rangle - \langle\hat{r}_i\rangle\langle\hat{r}_j\rangle \tag{77}$$

with $\hat{\mathbf{r}} = (\hat{x}_1, \cdots, \hat{x}_L, \hat{p}_1, \cdots, \hat{p}_L)$, the eigenvalues of the matrix $J\sigma$ are of the form $\pm i\lambda$ with $\lambda \geq \frac{1}{2}$. The matrix $J$ is

known as the symplectic matrix and encodes the commutation relations between the operators

$$[\hat{r}_i, \hat{r}_j] = iJ_{ij} \tag{78}$$

with $J = \begin{pmatrix} 0 & I \\ -I & 0 \end{pmatrix}$. For Gaussian pure states, all of the symplectic eigenvalues for the full correlation matrix have $\lambda = \frac{1}{2}$ which is the aforementioned normalization constraint.

For our numerical simulations of the dynamics of the correlation matrix we devised a method that explicitly preserves this normalization condition. To start we note that the pure state density matrix, used to calculate the correlation matrix, evolves according to

$$\hat{\rho}(t + \delta t) = e^{-i\hat{H}_{\text{eff}}\delta t}\hat{\rho}(t)e^{i\hat{H}_{\text{eff}}^\dagger \delta t} \tag{79}$$

The effective Hamiltonian can be naturally expressed in terms of it's hermitian and anti-hermitian parts

$$\hat{H}_{\text{eff}} = \hat{H} - i\gamma \sum_b \hat{O}_b^2$$
$$= \hat{H} - i\hat{D}.$$

Using the Baker-Campbell-Hausdorff formula, the evolution equation can be expressed

$$\hat{\rho}(t + \delta t) = e^{-i\hat{H}\delta t}e^{-\hat{D}\delta t}\hat{\rho}(t)e^{-\hat{D}\delta t}e^{i\hat{H}\delta t} + \mathcal{O}(\delta t^2). \tag{80}$$

For sufficiently small time step $\delta t$ the unitary and dissipative parts of the evolution can be treated separately. The effect of the unitary part of the dynamics on the correlation matrix is a simple well know result (linear symplectic transformation $S(\delta t)$) whilst the transformation of the correlation matrix under the dissipative part of the dynamics is not. This non-linear transformation $\mathcal{F}_{\delta t}$ can be derived exactly using a very similar method presented earlier resulting in a combined transformation of the form

$$\sigma(t + \delta t) = S(\delta t)\mathcal{F}_{\delta t}(\sigma(t))S(\delta t)^T \tag{81}$$

which when expanded to first order in $\delta t$ yields the correct Riccati equation whose general form has already been derived in [1, 2]. To the best of our knowledge we are the first to calculate the non-linear transformation $\mathcal{F}_{\delta t}(\sigma(t))$.

For the figures in the main text a time step $\delta t = 0.02$ was used.

The system was initialised in the following translationally invariant product state

$$|\psi(0)\rangle = \prod_j \exp\left(\frac{\alpha}{2}\hat{a}_j^\dagger \hat{a}_j^\dagger - \frac{\alpha^*}{2}\hat{a}_j \hat{a}_j\right)|\text{VAC}\rangle \tag{82}$$

with $\hat{a}_j|\text{VAC}\rangle = 0$. Here the creation/annihilation operators are the local harmonic oscillator ladder operators, $\hat{a}_j = \frac{1}{\sqrt{2}}(\hat{x}_j + i\hat{p}_j)$. For the figures in the main text we set $\alpha = 0.1$.

For initial states of this form, it can be shown that the effects of inter-band correlations on the entanglement growth are sub-leading which is the reason as to why the quasiparticle picture described in the main text captures the leading order behaviour so well.

The effective Hamiltonian is parameterised by mass $m$, measurement range $R$ and the measurement strength $\gamma$. Throughout this work we have fixed $m = 1$ however this does not have any qualitative effect on the results. For example the usual distinction between $m = 0, m \neq 0$ does not apply here since an effect of the measurements (non-hermitian part of $\hat{H}_{\text{eff}}$) is to create a gap in the real part of the quasiparticle spectrum if there is not already one.

## FINITE SIZE EFFECTS

In a finite system the entanglement does not grow indefinitely. The quasiparticle picture predicts that the entanglement will exhibit 'revivals' since the EPR pairs will reunite at multiples of the time $t_L = \frac{L}{2|v_j(k)|}$. We find that the quasiparticle picture has excellent agreement with the numerics (Fig. 3) where we have included a higher order term $(\delta k^2)$ in our expansion of the quasiparticle bandstructure in order to capture dispersion effects. The inclusion of this higher order term leads to the pronounced curvature of the revival peaks/troughs.

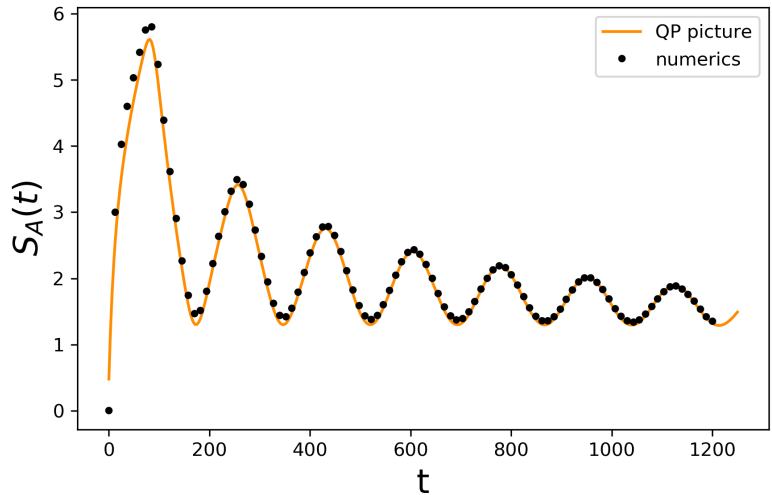

FIG. 3. Half chain entanglement for a system of size $L = 400$ for measurements of range 4 and strength $\gamma = 1$. We see perfect agreement between the numerical simulations and the quasiparticle picture prediction. The $t^{-1/2}$ decay was observed numerically by plotting the decay of the entanglement revival peaks with time.

## EFFECT OF INTER-BAND CORRELATIONS

In the main text we outlined the quasiparticle picture formalism for a multi-band model with no inter-band correlations. The additional correlations between the bands complicates the quasiparticle picture slightly and has already been studied in the context of free fermions under unitary dynamics [3]. Here they find that the unitary dynamics still give rise to ballistic growth of entanglement entropy. In our case there is also additional dissipative dynamics that can be treated separately. The correlation matrix has the general form for small $k$

$$\sigma(k,t) = U^\dagger(k,t) f(k^2 t) U(k,t) \tag{83}$$

where $U(k,t) = e^{-i\epsilon_k C t}$ is a unitary matrix and $f(k^2 t)$ is a matrix valued function. From this form we would still expect diffusive growth of entanglement entropy, even in the presence of inter-band correlations.

We provide some numerics (Fig. 4) to show that the qualitative behaviour of the entanglement growth does not change when there are inter-band correlation contributing at leading order. We consider the following initial state

$$|\psi(0)\rangle = \prod_b \exp\left( \frac{1}{2} \sum_j \left[ \alpha_j \hat{a}^\dagger_{b,j} \hat{a}^\dagger_{b,j} - \alpha_j^* \hat{a}_{b,j} \hat{a}_{b,j} \right] \right) |\text{VAC}\rangle \tag{84}$$

where we have now introduced two indices for the local operators with $b$ representing the unit cell and $j$ representing the position within the unit cell. States of this form with $\alpha_j$ distinct have the desired inter-band correlations contributing to the entanglement at leading order.

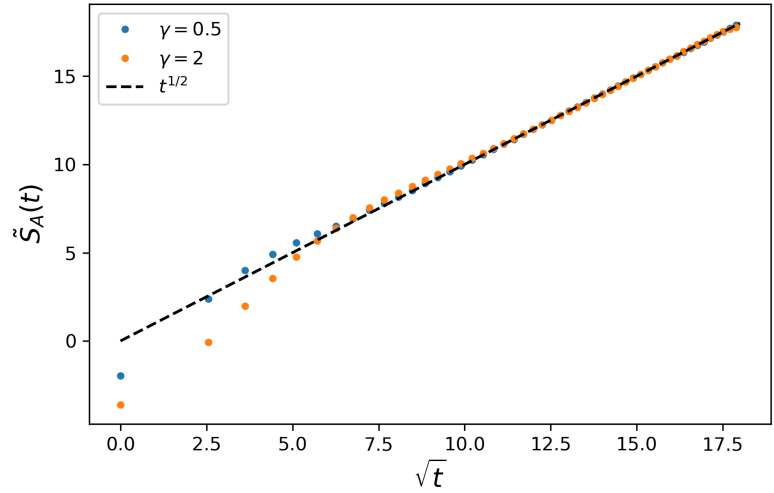

FIG. 4. Half chain entanglement growth for a system of size $L = 1600$ for measurements of range 4 where there are inter-band correlations at leading order. The entropy curves $S_A(t)$ are fitted to a curve $f(t) = a + b\sqrt{t}$ from which we define $\tilde{S} = \frac{S-a}{b}$. We see that the asymptotic entanglement growth is still diffusive even in the presence of these correlations.



\end{document}